\documentclass[preprint,showpacs, preprintnumbers, pra,amsmath,amssymb]{revtex4}

\usepackage{graphicx}
\usepackage{dcolumn}
\usepackage{bm}

\begin{document}

\title{Classical approach to the graph isomorphism problem using quantum walks}

\author{Brendan L Douglas}
\email{brendan@physics.uwa.edu.au}
\author{Jingbo B Wang}

\affiliation{School of Physics, The University of Western Australia, 6009, Perth, Australia.}

\begin{abstract}
Given the extensive application of classical random walks to classical algorithms in a variety of fields, their quantum analogue in quantum walks is expected to provide a fruitful source of quantum algorithms. So far, however, such algorithms have been scarce. In this work, we enumerate some important differences between quantum and classical walks, leading to their markedly different properties. We show that for many practical purposes, the implementation of quantum walks can be efficiently achieved using a classical computer. We then develop both classical and quantum graph isomorphism algorithms based on discrete-time quantum walks. We show that they are effective in identifying isomorphism classes of large databases of graphs, in particular groups of strongly regular graphs. We consider this approach to represent a promising candidate for an efficient solution to the graph isomorphism problem, and believe that similar methods employing quantum walks, or derivatives of these walks, may prove beneficial in constructing other algorithms for a variety of purposes.
\end{abstract}


\maketitle

\section{Introduction}

The question of determining whether two given structures (for instance, algebraic or combinatorial) are isomorphic has been a long-standing open problem in mathematics. A range of structures can be efficiently (i.e. in polynomial time) encoded by graphs \cite{Miller}, so for these it is sufficient to solve this problem for graphs. Considerable and continuing effort has been devoted to the graph isomorphism (GI) problem, due to both the variety of practical applications, and its relationship to questions of computational complexity.

Efficient GI algorithms exist for certain restricted classes of graphs, such as trees \cite{trees}, planar graphs \cite{Hopcroft}, and graphs with bounded valence \cite{Luks2}. Indeed, for many practical applications, the GI problem can be viewed as `easy', in that algorithms exist, such as Brendan Mackay's `Nauty' algorithm \cite{nauty}, which can distinguish pairs of graphs taken from many classes of graphs efficiently. However as yet, there is no known algorithm with polynomial bounds to solve GI for general graphs. Currently, the best known general GI algorithms scale with $O(e^{\sqrt{n \log{n}}})$ \cite{scaling, scale2, scale3}. In fact, the exact complexity status of GI is unknown and generates much of the interest surrounding the problem. No NP-completeness proof has been found, and indeed it seems likely that the GI problem is not NP-complete, in part because the corresponding testing and counting problems are polynomial-time equivalent, unlike the apparent case for all other known NP-compete problems. In these cases, the counting problem seems to be much harder, although of course this cannot be proven without solving the $P = NP$ problem.

Just as isomorphism testing of certain structures is equivalent to graph isomorphism, there are various other problems with equivalent difficulty. Problems such as finding the automorphism group of a graph \cite{aut}, and several subsets of GI, such as isomorphism of regular graphs, connected graphs, undirected graphs, and graphs of diameter 2 and girth 4 are all examples of problems equivalent to GI in general \cite{GIold}. In other words, solving GI for any of the above classes of graphs would provide an efficient method for solving GI in general.

In this work, the emphasis is on polynomial versus non-polynomial time computability of GI. Although we detail the specific scaling of the proposed algorithms to follow, we recognize that their current implementation may not be optimized, and it is likely that variations based on the same general theme could be developed that scale with lower order polynomials. For now, however, the important theoretical direction is towards establishing a polynomially scaling GI algorithm.

In this paper we present an algorithm utilizing the action of quantum walks on graphs to distinguish given pairs of graphs. Firstly, we introduce the concept of a quantum walk, and relate it to the more familiar classical random walk. We then discuss possible efficient, classical implementations of quantum walks on general graphs. A detailed discussion of a resulting GI algorithm follows, including scaling arguments, and results from testing against various sets of graphs. Finally, the corresponding quantum GI algorithm is discussed.

\section{Quantum Walks}
\label{qw}

There has been considerable interest in quantum walks recently, with a variety of papers published regarding their properties \cite{prop1, prop2}, uses \cite{uses, uses2}, and possible physical implementations \cite{implementation}. They are motivated by notions of quantum systems, and indeed can be thought of as the quantum analogue of simple classical random walks.

Consider a classical random walk on a line. In its simplest form, it consists of flipping a coin, moving either left or right depending on the outcome of the coin flip, then repeating these steps. The coin provides a means of randomizing the chosen direction of propagation. For example, consider an unbiased discrete walk on the integers, in which direct walking is allowed only between consecutive integers. If the walk is started at the origin, then after one step the walker would be at $-1$ or $1$ with equal probability, and after two steps, at $0$ with probability $0.5$, and $-2$ or $2$ with probability $0.25$.

For our purposes, we identify three important differences in the definition of a quantum walk. Firstly, since it is defined in terms of a physically implementable quantum system, both the coin and shifting operations are constrained to be unitary. In addition, all possible paths of the walk are sampled simultaneously - in other words, we can think of the evolution of the walk as being the evolution of the probability amplitude distribution, rather than involving a walk along a distinct path. Finally, as a result of the previous two requirements, quantum walks have memory, in that the outcome of the `coin flip' depends on outcome of previous coin flips.

\begin{figure}[Hh*tb]
\includegraphics[width=5.5cm]{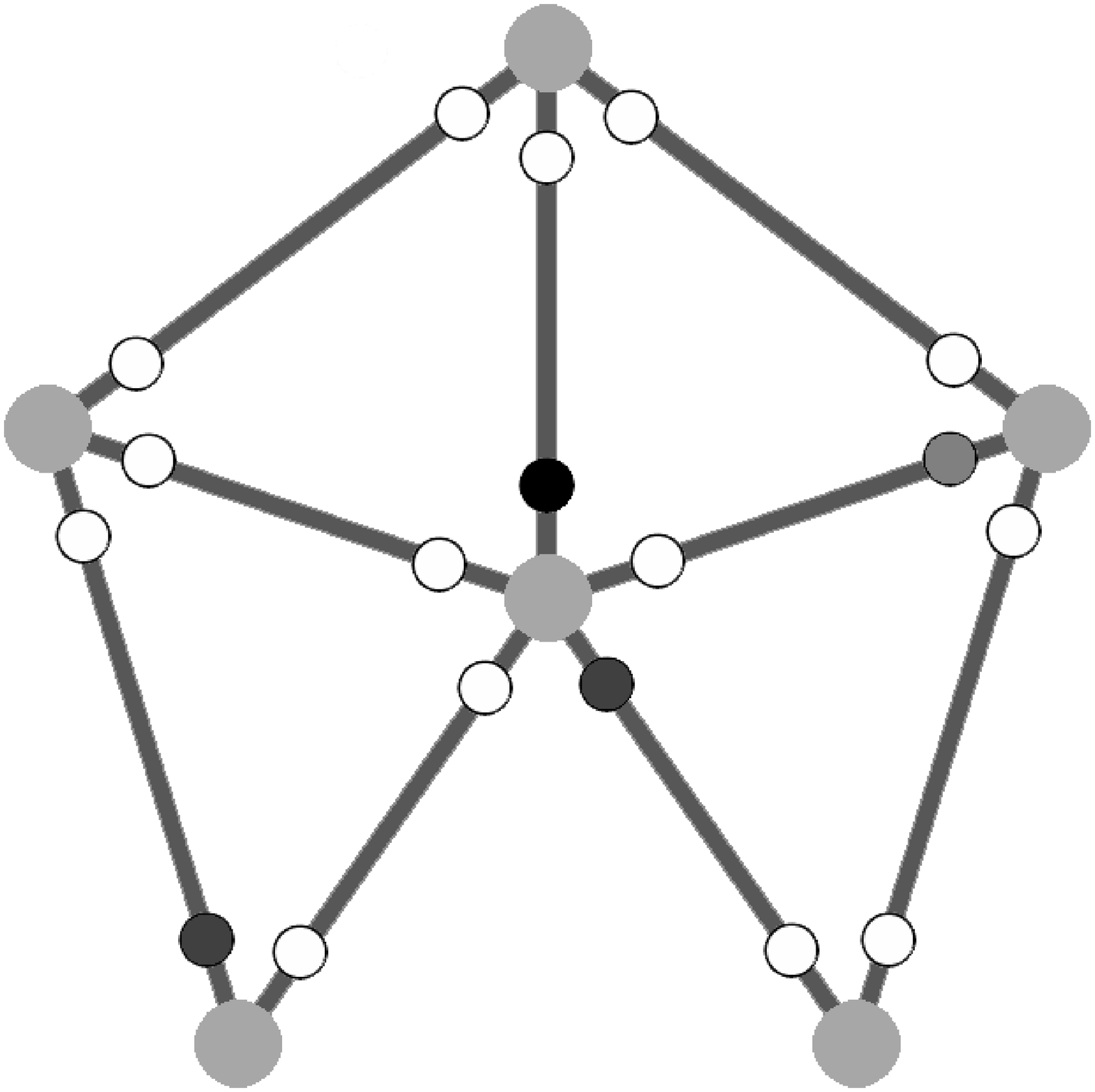}
\caption{\label{subnode} A sample graph of six nodes is shown, split into sub-nodes. Each edge is associated with two sub-nodes.}
\end{figure}

In fact, quantum walks in their present form have a one-step memory. This is achieved by increasing the state space of the walk, splitting each node of the position space into a group of $d$ sub-nodes (or `coin positions'), as illustrated in Figure \ref{subnode}, where $d$ is the number of outgoing `edges' from the node. In this case we are assuming that the walk is associated with some geometric structure containing edges, in which walking occurs only along these edges. The coin operator acts differently on each sub-node, thus giving the walk its memory.

\begin{figure}[htb]
\includegraphics[width=8.5cm]{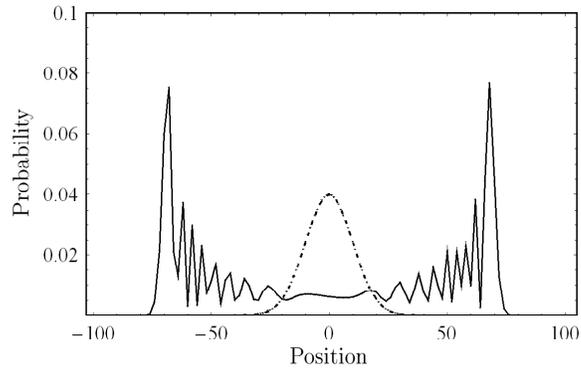}
\caption{\label{probdist} Probability distributions for unbiased quantum (solid) and classical (dashed) walks along a line.}
\end{figure}

Much of the interest in quantum walks has been based around their markedly different spreading characteristics, compared to those from simple classical random walks. Unbiased quantum walks along a line result in spreading with visible interference patterns such as that of Figure \ref{probdist} (solid line), whereas simple classical random walks give rise to a Gaussian distribution (dashed line in Figure \ref{probdist}). To investigate what causes these differences in propagation, we considered distributions resulting from altered quantum walks, in which one or more of the three requirements above were relaxed. We found that both the memory of the walk and the simultaneous sampling of all paths are needed to produce similar probability distributions to Figure \ref{probdist} (solid line). The combination of these two properties results in interference/interactions between different paths of the walk, not seen in classical random walks and certainly not possible with Markovian processes. This interaction between paths is the key difference between classical random walks and quantum walks. The third requirement, that of unitarity, is not strictly necessary to produce such probability distributions, and can be relaxed for classical implementations of these walks. It is however, necessary for a closed quantum system, and is an implicit property of any quantum implementation.

For the remainder of this paper we will only be utilizing the probability amplitude distributions associated with these walks. This removes any random aspect to both walks, as the evolution of the probability distributions is in both cases a deterministic process. It also removes one of the major nominal differences between the definitions of quantum and classical random walks described above, with the classical random walk now also involving sampling of all possible paths simultaneously. However, though simultaneous, the sampling of different paths proceeds independently, unlike the quantum walk, due to the lack of memory between steps.

Quantum walks on graphs are defined as above, with the shifting operator defined intuitively in terms of the edge set of the graph, so that direct transitions are allowed only between vertices connected by an edge. Note that each vertex, or node, can be thought of as having a group of sub-nodes (or coin positions), each associated with an outgoing edge. This means that each edge is now associated with two coin positions, one from each vertex connected by the edge. In effect, this describes a directed graph, as each end of an edge is defined, and can be manipulated, separately.

For a more formal mathematical definition of quantum walks along graphs, consider an undirected graph $G(V,E)$, characterized by a set of vertices (or nodes) $V=\left\{v_1,v_2,v_3,\ldots\right\}$, together with a set of edges $E=\left\{(v_i,v_k),(v_k,v_l),\ldots\right\}$, being unordered pairs connecting the vertices. A step of the walk is composed of both a coin operation and a shifting operation (in either order). Associated with each node $v_i\in V$ with valency $d_i$ is a group of $d_i$ sub-nodes. Each sub-node is in turn associated with a directed edge outgoing from $v_i$, as described above. The shifting operator then acts on the extended position space, spanned by these sets of sub-nodes. Its action consists of swapping the probability amplitudes of the pair of sub-nodes associated with each edge. For example, representing the amplitude corresponding to the di-edge connecting $v_i$ to $v_j$ as $(\stackrel{\longrightarrow}{v_i,v_j})$, and the amplitude corresponding to the di-edge in the opposite direction, connecting $v_j$ to $v_i$, as $(\stackrel{\longrightarrow}{v_j,v_i})$, the action of the shifting operator is defined by $$S (\stackrel{\longrightarrow}{v_i, v_j}) =( \stackrel{\longrightarrow}{v_j,v_i}).$$

The coin operator acts on the same extended position space, and its action consists of independently mixing the probability amplitudes associated with each group of sub-nodes of a given node. For example, given an undirected graph $G(V,E)$ with $n$ vertices and $k$ edges (i.e. $\left|V\right| = n$ and $\left|E\right| = k$), there are $2k$ sub-nodes, or coin positions, hence the position space of the walk can be represented by a $2k$ column matrix. Then the shifting operator ($S$) can be represented as a ($2k \times 2k$) permutation matrix, and the coin operator ($C$) as a unitary ($2k \times 2k$) block diagonal matrix, with each block representing a group of sub-nodes associated with a node. Each step of the walk is then represented by a unitary ($2k \times 2k$) matrix $U$, where $U = S.C$, acting on the state space.

Note that quantum walks involve unitary evolution and are hence reversible. Unlike classical random walks, they do not approach a stationary probability distribution. Indeed, for graphs with a high degree of regularity or large automorphism group, the resulting probability distributions often exhibit periodic behaviour.

Observations of probability distributions resulting from quantum walks along various graphs showed a strong relationship between the complexity of the probability amplitude distribution, and the size of the automorphism group. Moreover, since the amplitude distributions obtained showed a high sensitivity to minor changes in the graph, it seemed likely that properties of the probability distribution might serve as an effective tool for producing a graph certificate that would enable the topological uniqueness of a graph to be established.

These observations provided the motivation behind looking into a possible GI algorithm based on quantum walks. Initially, this was with a view to quantum systems (and hence a quantum algorithm) only. However, while constructing the algorithm it became evident that for the purposes required here, quantum walks can be implemented efficiently on a classical computer.

Consider the situation above, with a quantum walk along a simple, undirected graph with $n$ nodes and $k$ edges. There are exactly $2k$ coin positions, where $k \le n^2$. Matrix multiplication between two matrices of size ($n \times n$) requires $O(n^3)$ computational time. Each step of the walk involves applying the shifting and coin operations to the state vector, of length $2k$. Then given shifting and coin operators both described by ($2k \times 2k$) matrices, each step of the walk can be implemented on a classical computer in a time that scales with O($k^3$), being at most $O(n^6)$. In fact the shifting operator can be implemented much faster than $O(n^6)$. A more straightforward implementation than matrix multiplication, involving directly rearranging the elements of the state space, has an upper bound scaling of $O(n^4)$. In this case, the permutation of the state space corresponding to the shifting operator is implemented by sequentially mapping between individual states, via a temporary section of computer memory. Similarly, the coin operator, a block diagonal matrix containing $n$ blocks each of size at most $n$, can also be implemented with an upper bound scaling of $O(n^4)$. Hence quantum walks on graphs, in the form described above, are efficiently implementable on a classical computer, with the computational time required to simulate a step of the walk being on the order of $O(n^4)$.

For the purposes of a GI algorithm, the coin operator must not be biased with respect to the labeling of the vertices. In particular, this means that each block (or `coin') of the coin matrix must be symmetric, with the form
\begin{equation}
C_{i,j} = \left\{ \begin{array}{rcl}
a && i = j\\
b && i \ne j\\
\end{array}
\quad\textrm{where}\; a,b \ne 0.\right.
\end{equation}

Since each coin must also be unitary, the Grover coin, having the definition $C_{i,j} = \frac{2}{d} - \delta_{i,j}$ is the only purely real, non-trivial coin satisfying the constraint that symmetry with respect to labeling is preserved. Similarly, the shifting operator must not be biased with respect to labeling. Hence its action can only consist of either the identity operation, or shifting along the edge with which the corresponding coin position is associated. As a consequence, the shifting operator applied twice is simply the identity operator.

\section{Proposed GI Algorithm}
\label{algorithm}

A simple GI algorithm directly employing quantum walks could consist of starting in an equal superposition of all states, evolving a quantum walk along two graphs for some fixed number of steps, then comparing the two sets of probability amplitude distributions obtained. If the probability amplitude at one node of one graph had no corresponding match for any node of the other graph, the graphs would be necessarily distinct. Consider two graphs in which each node on one graph has the same number of cycles of every length as at least one node on the other graph. We will term such graphs `walk-equivalent'. It is clear that the method above distinguihes graphs which are not walk-equivalent. Indeed, such a naive method is quite effective at distinguishing most graphs. However, the method fails to distinguish very similar graphs, including some vertex transitive graphs, and all strongly regular graphs with the same parameters.

Similar methods have been independently attempted in previous work \cite{shiau, severini}. Shiau et. al. performed single-particle quantum walks on closed graphs, but they concluded that such walks fail to identify non-isomorphic strongly regular graphs \cite{shiau}. Emms et. al. introduced a new matrix representation inspired by quantum walks \cite{severini}. This representation appears to yield different spectra for sets of non-isomorphic strongly regular graphs, but it fails to distinguish other regular graphs, and general non-regular graphs.

Note that methods which attempt to solve GI in general tend to break down for certain difficult classes of graphs, most notably strongly regular graphs, pairs of which are commonly used as test cases for proposed GI algorithms. A strongly regular graph (SRG) with parameters $(n,d,\lambda,\mu)$ is a $d$-regular graph of order $n$, such that all pairs of adjacent nodes have exactly $\lambda$ common neighbors, and all pairs of non-adjacent nodes have exactly $\mu$ common neighbors.

So applied in their simplest manner, quantum walks cannot distinguish certain classes of graphs, including SRG's with the same parameters. However we have significantly more freedom available in the construction of such an algorithm. In this case, we can add certain inhomogeneities into the graphs, with the aim of breaking the symmetry with respect to the walks that exists between them. Such inhomogeneities could take a variety of forms, such as self loops or gadgets added to a node, extra nodes added along an edge, or phase additions to nodes or edges. Of these forms, phase represents a particularly elegant addition to such walks, in that the connectivity of the graph is left unaltered, and the definition of quantum walks allows for phase additions to nodes or edges (directed or undirected) without significant extra computational costs. Phase can be added equivalently to the coin operator (added to blocks of the coin matrix) or the shifting operator (added to directed edges), or as a separate third component of a step of the walk (added directly to the state space). For example, a $\phi$-phase can be added to a di-edge $(\stackrel{\longrightarrow}{v_i, v_j})$ by altering the action of the shifting operator from:
$$S (\stackrel{\longrightarrow}{v_i, v_j}) = (\stackrel{\longrightarrow}{v_j,v_i}) \textrm{, $\;$ to $\;$} S (\stackrel{\longrightarrow}{v_i, v_j}) = e^{i \phi} (\stackrel{\longrightarrow}{v_j,v_i}).$$

Although pairs of `walk-equivalent' graphs are often not distinguished using simple quantum walks, if phase additions are made to two or more nodes, cycles originating from, or involving, these nodes will also interact. The connectivity, or geometric relationship between such paths then becomes important, and affects the evolution of the walks.

\begin{figure}[htb]
\includegraphics[width=5.5cm]{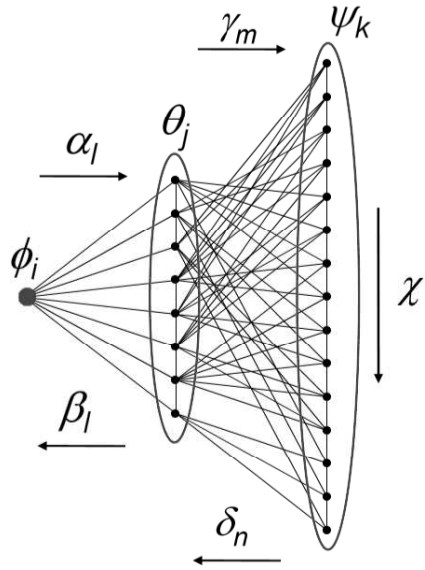}
\caption{\label{ranked} A given graph of diameter 2, showing all possible groups of phase additions, unbiased with respect to labeling, after fixing the labeling of a single node.}
\end{figure}

Any GI algorithm as a whole must not be biased with respect to the labeling of the vertices, since we require identical results to be produced from all possible permutations of a graph. So if phase is added about certain nodes (termed `reference nodes'), we need to cycle over all possible such selections within the algorithm. Given some arbitrary vertex of a graph as a reference node, the labeling of this node is considered fixed, however to preserve the arbitrary labeling of all other nodes, phase additions can only be made to groups of nodes defined solely in terms of this reference node. The only such groups are the groups of vertices at each fixed distance from the chosen vertex, and the sets of edges mapping both within and between each of these groups, shown in Figure \ref{ranked}. Explicitly, for a chosen vertex $v_i$ of a graph $G(V,E)$, these are the sets of vertices:
$$ V_i^a \in \textrm{V, } V_i^a = \left\{v \in \textrm{V }: d(v_i,v) = a\right\},  $$
for any constant $a$, where $d(v,v_i)$ is the distance between $v$ and $v_i$, and the sets of edges: 
$$ E_i^b \in \textrm{E, } E_i^b = \left\{(v_j,v_k) \in \textrm{E }: min\left[d(v_i,v_j),d(v_i,v_k)\right] = b\right\},  $$
for any constant $b$. Distinct phase additions can be made to nodes or directed edges within each of these divisions, consistently with respect to labeling. If two or more reference nodes are selected, further divisions can be made, consisting of nodes or edges connecting these nodes, or at some fixed distance from two or more of them.

Given two isomorphic graphs, for each possible set of chosen reference nodes on one graph, there must be at least one corresponding set of reference nodes on the other graph yielding an identical probability amplitude distribution. The aim of a GI algorithm is then to implement a scheme of phase additions in which, for all non-isomorphic pairs of graphs, there are efficiently measurable differences between the resulting probability distributions.

\subsection{Algorithm}

Based on this phase addition scheme, we propose the following GI algorithm.

We are given a pair of graphs $A$ and $B$. For each graph $G(V,E)$ with order $n$, we initialize a quantum walk, starting in an equal superposition of all states. We then (arbitrarily) choose two nodes $v_1$ and $v_2$ on the graph, termed `reference nodes'. For each of these two nodes, consider the groups:
\begin{eqnarray}
G_0 &=& V_i^0, \quad i\textrm{ = 1 or 2} \nonumber\\
G_1 &=& V_i^1, \textrm{and} \\
G_2 &=& V / (G_0 \cup G_1), \nonumber
\end{eqnarray}
defined only in terms of the reference nodes. These groups are the reference node itself ($G_0$) and the groups of adjacent nodes ($G_1$) and non-adjacent nodes ($G_2$). Phase additions are made at each step of the walk to $G_0$, the edges connecting $G_1$ to $G_2$, and to the edges connecting $G_1$ to $G_0$. The walk is evolved for a number of steps with an upper limit of $2n$, to ensure that the walk samples all areas of the graph, with any differences in amplitude propagating back to the reference nodes. At each step of the walk, the probability amplitude associated with the node $v_1$ is recorded. The process is then repeated for all $n^2$ possible choices of $v_1$ and $v_2$, producing $n^2$ sets of probability amplitudes.

In order to establish a `baseline' measure of the symmetry of each graph individually with respect to these walks, the sets are compared pairwise. In other words, each set of $n$ probability amplitudes, corresponding to some pair of reference nodes, is compared to every other set, building up a comparison table with $n^4$ elements. If the set resulting from one pair of reference nodes $v_i$ and $v_j$ matches that resulting from another pair $v_k$ and $v_l$, we write a ``0'' on the comparison table, in the position $(i,j,k,l)$. Otherwise, if the sets do not match, this is recorded as a ``1''.

For the two given graphs, we then obtain two tables, labeled (A-A$'$) and (B-B$'$), representing the comparison of each graph to its own permutation. A third table (A-B), also containing $n^4$ elements, is constructed by pairwise comparison of the sets of probability amplitudes from one graph to those from the other graph. These tables provide a measure of the relative amplitude distributions resulting from walks along the two graphs. In effect, the (A-A$'$) table represents a comparison of graph A to itself (or equivalently to a permutation of itself), whereas the (A-B$'$) table represents a direct comparison between the amplitude distributions resulting from graphs A and B.

Comparing the amplitudes of the reference nodes (i.e. partial states) is for our purposes equivalent to comparison of the entire distributions, since differences in parts of the distributions propagate through to all nodes with repeated application of the walks. In effect, repeated steps of the walk results in a recursive partitioning of the vertex set based on the relative connectivity of each node from the pair of reference nodes. In other words, differences between the overall states after some number of steps (manifested as differences between the amplitudes of some small subset of nodes in each graph) are propagated throughout the entire distribution after repeated steps of the walk, and in particular propagated to the reference nodes, which are then measured.

If the comparison table (A-B) is different to (A-A$'$) or (B-B$'$) (beyond a permutation), the two graphs A and B are necessarily non-isomorphic. For the purposes of the algorithm, the total number of ``1''s in the three comparison tables (A-A$'$), (B-B$'$) and (A-B) are compared. If these totals differ, the graphs A and B are again necessarily different. The important question is whether the converse is true - whether equal sets of probability amplitude distributions resulting in equal comparison tables imply that the graphs \textit{are} isomorphic. Computational testing against databases of graphs, detailed in Section \ref{testing}, supports this possibility, and attempts to establish a formal proof are currently in progress.

\subsection{Discussion of algorithm}

The use of two reference nodes is a vital aspect of this algorithm. Addition of phase about only one reference node fails to identify sufficient connectivity information. For instance, the properties of strongly regular graphs (namely that they are regular, all pairs of adjacent nodes have the same number of common neighbors, and all pairs of non-adjacent nodes have the same number of common neighbors) mean that they cannot be distinguished through the use of a single reference node, since the propagation of phase from this node will be identical for SRG's with the same parameters. Specifically, the number of paths of any given length between two points at some fixed distance is identical, and hence the propagation of phase throughout the graph, eminating from a single reference node is also identical.

However given two or more reference points, the complex amplitudes associated with paths originating from both points will interact, leading to differences in the resulting amplitude distributions of the walks. Taking further steps of the walks recursively partitions the vertex set, based on the relative connectivity of the nodes. Essentially, differences occuring within the probability amplitude distributions arising from quantum walks along two graphs will propagation through the distribution. For instance, if the amplitude at node $x$ is different after some number of steps of the walk, all nodes connected to $x$ will have different amplitudes after taking further steps of the walk. Expressing the veracity of the above GI algorithm in an alternate form, it requires that all isomorphism classes of graphs are distinguished based on the connectivity of nodes relative to all possible pairs of reference points.

Given differences in the amplitude distributions resulting from walks along a pair of graphs, these differences will appear in the comparison tables produced by the algorithm, as ``1''s in positions where the ``0''s (matches) would otherwise occur when comparing one of the graphs to a permutation of itself.

\begin{figure}[htb]
\includegraphics[width=8.5cm]{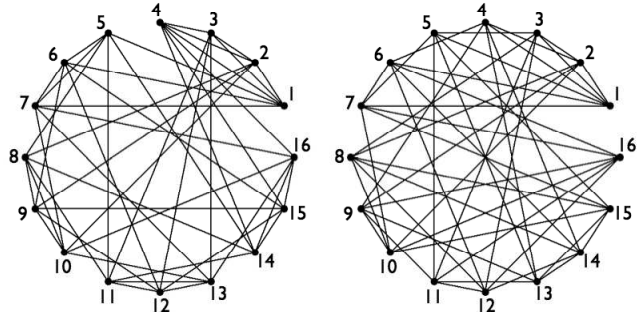}
\caption{\label{srgs16} The two strongly regular graphs with parameters (16,6,2,2).}
\end{figure}

\begin{figure}[htb]
\includegraphics[width=8.5cm]{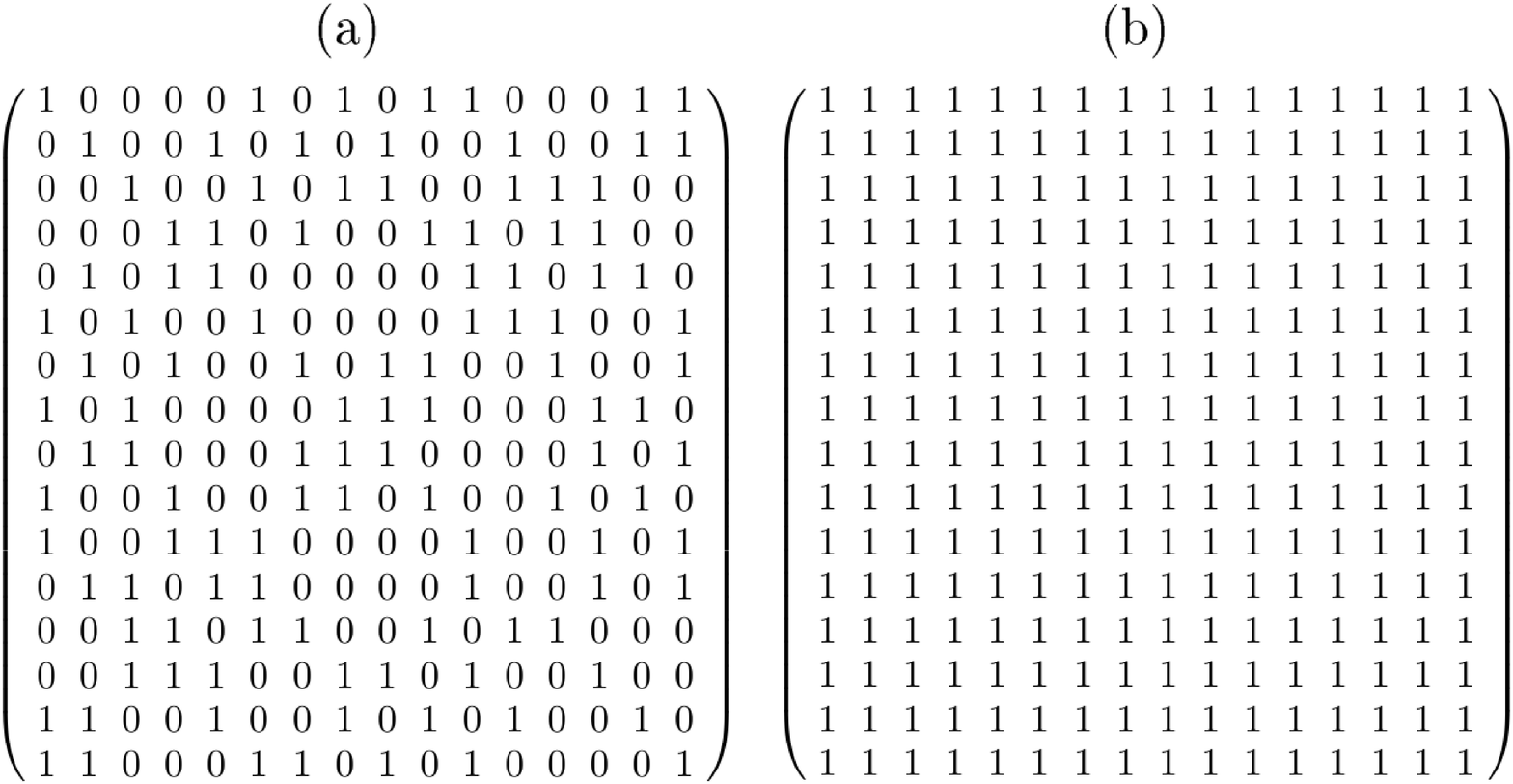}
\caption{\label{comp} Samples of the comparison tables resulting from the two strongly regular graphs of figure \ref{srgs16}. (a) and (b) represent the (A-A$'$) and (A-B) comparison tables respectively, from positions (1,2,*,*). If the graphs were isomorphic, these tables would necessarily contain the same number of ``1"s and ``0"s.}
\end{figure}

For example, consider the two strongly regular graphs with parameters (16,6,2,2), shown in Figure \ref{srgs16}. Running through the above steps, we find that for the first graph, adding phase about nodes 1 and 2, the amplitude distribution measured at node 1 matches that obtained for 144 other pairs of nodes of this graph, shown in the comparison table of Figure \ref{comp}(a). However, it does not match the amplitudes obtained from any of the other pairs from the second graph, as shown in the comparison table of Figure \ref{comp}(b). If these two graphs were isomorphic this total number of matches would be necessarily identical. Hence, the graphs are distinguished.

\subsection{Complexity of algorithm}

The complexity of this algorithm can be ascertained by first noting (from Section \ref{qw}) that the computational time required to implement a step of a quantum walk along a graph of order $n$ scales with an upper bound of $O(n^4)$. Constructing the sets of probability amplitudes requires taking $n$ steps of the walk for each of the $n^2$ pairs of phase nodes, hence requiring a total computational time of $O(n^7)$. An upper bound of $O(n^4)$ comparisons are then made between elements of these sets of amplitudes. Hence the algorithm in its present form scales with $O(n^7+n^4) = O(n^7)$.

This complexity calculation, however, does not take into account the possible precision required for the probability amplitude. It may be possible that after the walk is evolved for the required number of steps, the differences in probability amplitudes between non-isomorphic graphs are exponentially small relative to the graph size. Since the information that is required when comparing a pair of graphs is the relative values of the probability amplitudes, and not their absolute values, it is possible that this problem can be circumvented (at least in the classical case), given appropriate alterations to the algorithm. This is the subject of further study however, and will be left as an open problem at this stage.

\subsection{Addendum to the algorithm - finding an isomorphism}

Since the problems of testing for isomorphic graphs, and finding a specific mapping (isomorphism) between isomorphic graphs are Turing equivalent \cite{aut}, we can extend the algorithm described above to provide an efficient (polynomial time) method to either find an isomorphism between graphs which are not distinguished and hence prove they are isomorphic, or prove that the algorithm does not solve GI.

Specifically, consider two graphs A and B which are not distinguished by the above algorithm. We choose a node of A, and fix its labeling, for instance by adding a self loop or gadget to the node. We repeat this for one node of B, fixing its labeling in the same way, and then compare the graphs using the above algorithm. We then repeat this method, cycling over all $n^2$ combinations of single nodes in A and B. If the two graphs are isomorphic, isomorphisms between them will be manifested as matches between the graphs. 

Hence any mapping compiled from such matches represents an isomorphism between the graphs. This mapping can then be applied to the adjacency matrices, and easily checked. If it does not represent a mapping between the adjacency matrices of the two graphs, then the algorithm cannot distinguish the graphs down to their isomorphism classes. The benefit of this addendum to the algorithm is that the end result of comparing a pair of graphs (which is arrived at in polynomial time) must be one of the following: definite knowledge that the graphs are not isomorphic, definite knowledge that the graphs are isomorphic, or a proof that the algorithm does not solve GI.

\section{Testing algorithm}
\label{testing}

We tested the above algorithm against a variety of graphs, listed in Tables \ref{table1} and \ref{table2}, taken from databases at \cite{data1,data2,data3}. A wide variety of classes of graphs were chosen, in order to demonstrate that the algorithm is not limited to a specific type of graph. In particular, a large number of strongly regular graphs (SRG's) were examined, as groups of SRG's with the same parameters are particularly similar, and often used as test cases for proposed GI algorithms. In addition, groups of projective planes were chosen as they again represent a particularly difficult class of graphs to distinguish. In each case, all graphs were compared pairwise.

In practice, most graphs do not need to be compared directly as described in the previous section. Instead, given two graphs, each graph can first be compared to its own permutation, with the total number of ``0"s recorded as above summed, resulting in a single integer. This represents the total number of matches obtained from pairwise comparisons of the probability amplitudes resulting from each pair of reference nodes. If this total differs between the two graphs, they are again necessarily distinct. Otherwise, they can then be compared directly. This presents the possibility of developing a succinct graph certificate using a similar method. Currently, this scheme only yields an incomplete certificate, in that the final integer obtained is not unique to the graph. However, it is still quite useful for pairwise comparisons of large groups of graphs. All graphs of each set in Tables \ref{table1} and \ref{table2} were first compared indirectly, via this incomplete graph certificate, greatly decreasing the required number of direct comparisons. For example, there are 32548 strongly regular graphs with parameters (36,15,6,6), requiring $~5.3 \times 10^8$ pairwise comparisons. Of these however, only $~2 \times 10^5$ pairs possessed the same certificate, and were then distinguished with direct comparisons. All non-isomorphic graphs tested were successfully distinguished.

\begin{table}[htbp]
\caption{\label{table1}Groups of SRG's tested (all graphs in each group were compared pairwise).}
\begin{ruledtabular}
\begin{tabular}{ll}
Parameters&Group size\\
\hline
(16,6,2,2)&2\\
(25,12,5,6)&15\\
(26,10,3,4)&10\\
(28,12,6,4)&4\\
(29,14,6,7)&41\\
(35,18,9,9)&227\\
(35,16,6,8)&3854\\
(36,14,4,6)&180\\
(36,15,6,6)&32548\\
(37,18,8,9)&6760\\
(40,12,2,4)&28\\
(45,12,3,3)&168\\
(64,18,2,6)&78\\
\end{tabular}
\end{ruledtabular}
\end{table}

\begin{table}[htbp]
\caption{\label{table2}Groups of other graphs tested.}
\begin{ruledtabular}
\begin{tabular}{lll}
Graph type&Order of graph&Group size\\
\hline
Eulerian&8&184\\
&9&1782\\
&10&33120\\
Cubic vertex-transitive&48&32\\
&60&26\\
&64&38\\
Planar&7&646\\
Tree&14&3159\\
Vertex-critical&10 ($\chi$ = 4)&2453\\
Edge-critical&12 ($\chi$ = 7)&395\\
Vertex-transitive&29&1182\\
Hypohamiltonian&26&2033\\
Projective planes of order 16&273&13\\
Projective planes of order 49&4902&200\\
\end{tabular}
\end{ruledtabular}
\end{table}


It is apparent that there is considerable freedom associated with the chosen phase additions in such a method. An alternate, simpler phase scheme could involve the addition of phase, at each step, to only the two chosen reference nodes. Even simpler would be to restrict this phase to a $\pi$-phase addition. In effect, restriction to $\pi$-phase additions means that we are no longer working with phases encompassing the one dimensional complex unit circle, but rather two points - the set $\left\{\pm 1 \right\}$. This simplified scheme was tested against all the graphs of tables 1 and 2, distinguishing all pairs expect one, a pair of SRG's with parameters (40,12,2,4). Specifically, these are the point graph of the generalized quadrangle GQ(3,3) and its dual (i.e. the point graph and line graph of GQ(3,3)). These graphs are particularly similar, being both distance transitive (and hence rank 3), and having the same size automorphism group. Given that they were an exception amongst all graphs tested, being the only non-isomorphic pair not distinguished with the simplest two-phase scheme, several other rank 3 graphs, together with additional generalized quadrangles were tested. In particular, GQ(5,5) and GQ(7,7) (having orders 156 and 400 respectively), together with their respective duals, were compared. Again, the simple two-phase scheme was not sufficient to distinguish them, with the original scheme, involving further phase additions, to edges, required. All other pairs of rank 3 graphs tested, ranging in order from 49 to 364, were distinguished using the simple two-phase scheme.

\section{Quantum implementation}

Up to this point only a classical implementation of a GI algorithm employing quantum walks has been discussed. However as the name suggests, quantum walks can be very efficiently implemented on a quantum computer, or some quantum system specifically designed to implement them. The advantage of a quantum system is the availability of a state space, or memory that grows exponentially with the number of qubits, allowing in particular for a quantum walk to be evolved more efficiently than on a classical computer. 
There are also disadvantages in the quantum implementation however. The complete knowledge of the evolution of the probability amplitude distribution is not available for a quantum system, and instead repeated measurements must be made to approximate the parts of the wavefunction of interest. Hence any algorithm developed would no longer be deterministic. In some ways the quantum implementation has less freedom in the possible properties of the walk, in that it is restricted to be a unitary process. Depending on the properties of the implementation used however, potentially useful expansions to the algorithm may be accompanied by significantly less, if any, additional costs, compared to those associated with a classical implementation.

\begin{figure}[htb]
\includegraphics[width=8.5cm]{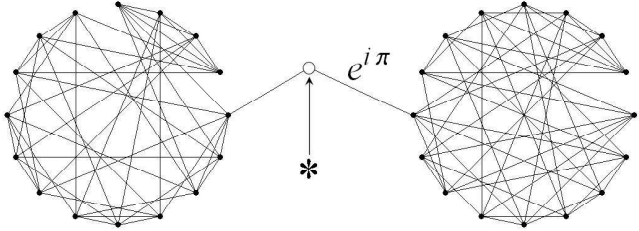}
\caption{\label{measurement} An illustration of the proposed quantum measurement scheme. The amplitude at two nodes is compared by connecting them via an additional node introduced into the system, at which the measurement is made.}
\end{figure}

Modifications must be made to this algorithm to implement it within the constraints of a quantum system. Specifically, since the probability distribution is not directly known, the measurement process requires some important changes. We can no longer simply record the probability amplitude at one of the phase nodes at each step. Rather than recording these amplitudes from each graph separately, then comparing them, the amplitudes from the two graphs are directly compared. After evolving the walk along each graph separately for some given number of steps, an additional node is introduced into the system, connecting the two nodes on each graph that are to be compared, as shown in Figure \ref{measurement}. The probability amplitudes in each of these two nodes are shifted into the connecting node, with a $\pi$-phase change applied to the wavepacket entering from one direction. If the connected vertices from each graph have identical probability amplitudes, the $\pi$-phase change will cause them to cancel out exactly, otherwise there will be some non-zero probability to find the walk at this connecting node.

The advantage of this method lies in the requirement of only one measurement, which is a direct comparison between the graphs. If the connected nodes are equivalent with respect to the walk, there will always be zero probability to measure the walk at the connecting node, hence the measurement itself will not alter the system. If they are not equivalent, the effects from the measurement will not matter, as all that is required is a zero or non-zero measurement. Note that such a measurement will still need to be repeated many times (or until a non-zero measurement is made) to achieve the desired level of accuracy. In particular, assume the connected nodes are not equivalent with respect to the walk, and the average probability of a non-zero measurement is $\frac{1}{p}$. Assuming this probability remains constant at each step \footnote{Fluctuations will occur, however they will increase rather than decrease the level of accuracy after a fixed time, providing this time is larger than the time scale of the fluctuations}, the probability $P(m)$ of a non-zero measurement after $m$ steps, is:
\begin{eqnarray}
P(m) &&= \sum_{i=0}^{m-1}\left[\frac{1}{p}\left(\frac{p-1}{p}\right)^i\right]\nonumber\\
&&\approx 1 - e^{-m/p} \textrm{ for large $p$}.
\end{eqnarray}
Assuming the walk on a graph of order $n$ is on average evenly distributed along all $n$ nodes, the average probability is simply $\frac{1}{n}$. In practice, the probability amplitude of the walks is often concentrated in small subsets of a graph. Accounting for this by taking the average probability at a given node to be $\frac{1}{n^{c}}$, for some constant $c$, we then obtain:
\begin{equation}
P(m) \approx 1 - e^{m/n^{c}},
\end{equation}
giving an error after $m$ measurements of approximately $e^{m/n^{c}}$. So to obtain any fixed accuracy, the number of required measurements scales with $O(\textrm{log}(n^c)) = O(\textrm{log}(n))$.

Considering the scaling of this quantum algorithm, we note that the two steps of obtaining the sets of amplitudes, and comparing these sets, can no longer be performed separately within the algorithm. Instead, for each pair of phase nodes on one graph, we must cycle over all $n^2$ possible pairs on the other graph, for each one comparing the amplitudes as described above, leading to $n^4$ possible combinations. $O(n)$ steps are required for each set of parameters. Assuming an efficient implementation of the quantum walks, the time to execute one step of the walk scales with $O(\textrm{log}(n))$. The complete algorithm would then scale with $O(n^5 \textrm{log}(n)^2) < O(n^6)$.

As in Section \ref{algorithm}, this complexity consideration is only correct if the differences between the amplitudes of non-isomorphic graphs are assumed to scale at worst polynomially with graph size. Otherwise the $\frac{1}{n^c}$ factor is no longer valid, and the complexity of the algorithm may no longer be polynomial in $n$.

Note that the actual space explored in the algorithm (via the walk along the graph) involves a memory with an upper bound of $n^2$, where $n$  is the number of vertices. Specifically the memory required scales with $2e \le n(n-1) $, where $e$ is the number of edges in the graph. Since there will only be exactly $2e$ states accessed by the walk, requiring in theory only $\textrm{log}(2e)$ qubits in a quantum implementation. This allows a more efficient implementation of quantum walks than that attainable with the classical method. Nonetheless, the polynomial growth (with graph size) of the state space of a quantum walk along a graph is the reason why quantum walks can also be simulated in polynomial time on a classical computer.

\section{Conclusion}

We hope that this work will aid in a deeper understanding of the properties of quantum walks and how they differ from simple classical random walks. For the purposes of evolving such walks within a polynomially growing state space, without the presence of noise or quantum measurements, we see that they are classically efficiently implementable. It is an interesting question as to which algorithms exist employing quantum walks that cannot be implemented classically.

With regards to the classical GI algorithm developed in this work, several open questions remain. The algorithm successively distinguished all non-isomorphic graphs tested, including groups of graphs traditionally considered to be `difficult' to distinguish. How to prove whether it can distinguish general graphs is the focus of further work. The general methods employed here may also prove beneficial in the construction of other algorithms. The measurement method utilized by the quantum GI algorithm is quite novel, and represents an effective means of extracting information from system wavefunctions without undesirable disturbances to the system.

Possible derivatives of quantum walks, for instance with multiple-step memories, may also provide fruitful areas of future study. Finally, the key property of quantum walks used here to distinguish graphs, namely the interactions between all possible paths through the state space, conceivably has a variety of other uses. For problems in which rapid sampling of the state space, and control of the interactions between paths through this space are desired, quantum walks provide promising candidates for their solution.

\begin{acknowledgments}
We would like to thank Gordon Royle for some fruitful discussions and very useful data. We thank iVEC for an intership grant, vital technical support and use of their SGI Altix supercomputer. We also thank A. Woods, C.H. Li, C. Praeger and J. Twamley for their interest in the work, and valuable discussions.
\end{acknowledgments}

\newpage

\end{document}